\documentclass[useAMS,usenatbib]{mn2e}

\usepackage{graphicx}
\usepackage{latexsym}
\usepackage{float}
\title{The 43GHz SiO maser in the circumstellar envelope of the AGB star R
  Cassiopeiae}   
 
\author[K. A. Assaf et. al.]{K. A. Assaf$^1$, P. J. Diamond$^{1,2}$,
A. M. S. Richards$^1$ and M. D. Gray$^1$\\ 
$^1$JBCA, Alan Turing Building, School of Physics and Astronomy, The
University of Manchester, Oxford Road, Manchester, M13 9PL, UK\\
$^2$CSIRO Astronomy and Space Sciences, PO Box 76, Epping, NSW 1710, Australia}

\begin{document}

%\date{Accepted 2010 December; Received 2010 December %14; in original
  %form 2010 October 11} 

\pagerange{\pageref{firstpage}--\pageref{lastpage}} \pubyear{2010}

\maketitle

\label{firstpage}
\begin{center}
\begin{abstract}

    We present multi-epoch, total intensity, high-resolution images of
43GHz, v=1, J=1-0 SiO maser emission toward the Mira variable R
Cas. In total we have 23 epochs of data for R Cas at approximate
monthly intervals over an optical pulsation phase range of $\phi=
0.158$ to $\phi= 1.78$. These maps show a ring-like distribution of
the maser features in a shell, which is assumed to be centred on the
star at a radius of $1.6 \to 2.3$ times the radius of star,
$R_\star$. It is clear from these images that the maser emission is
significantly extended around the star. At some epochs a faint outer
arc can be seen at 2.2 $R_\star$. The intensity of the emission waxes
and wanes during the stellar phase. Some maser features are seen
infalling as well as outflowing. We have made initial comparisons of
our data  with models by \citet{gray}.

\end{abstract}
\end{center}

\begin{keywords}
maser -- techniques: interferometric -- star: AGB stars and post-AGB --
star: individual: R Cas .
\end{keywords}

\section{Introduction}

   The Asymptotic Giant Branch (AGB) is the last evolutionary stage of
low- and intermediate-mass stars that are nuclear powered. This phase
of evolution is characterised by 
nuclear burning of hydrogen and helium in thin shells above the
electron degenerate core of carbon and oxygen. These stars are
very large (R$\sim$1 AU), very cool (2000-3000 K) and have abundant,
molecular, dusty winds (mass loss rate $10^{-7} - 10^{-5}$
\,M$_\odot$ \,yr $^{-1}$  \citep{herwing}.   

Any AGB star in the end will lose mass in the form of a slow wind at a
rate that will significantly affect the mass of the star. This
produces a circumstellar envelope of escaping dust and gas
particles. When the envelope becomes separated from the star, the
mass-loss sometimes ceases in a short period of time and the
temperature drops as the wind flows away from the star, from about
3130 K at the photosphere to $\approx$ 740 K at 5.5R$_{\star}$
\citep{gray}; cooling takes place by line radiation from various
molecules, especially H$_2$O. Some molecular species are formed in an
equilibrium process deep in the atmosphere and are destroyed in the
outer parts of the outflow by interstellar ultraviolet radiation
(H$_2$, CO, H$_2$O).  Other species, for example SiO, are depleted due
to condensation on dust particles at a few stellar radii
\citep{habing}. The observations show that almost all AGB stars are
pulsating variables, the pulsations produce waves which steepen up into shock
waves in the large density gradient of the stellar atmosphere. These
pulsational shocks are well-known to influence the dust formation zone
\citep{freytag}.  Stellar pulsations cause waves which travel into the
thinner outer layers of the atmosphere and develop into shocks with
temperature and density variations strong enough to trigger
intermittent dust formation. Large opacities allow the dust grains to
capture the stellar radiation efficiently, leading to an outward
acceleration of the dust particles which is transmitted to the gas by
collisions, causing a slow but dense outflow from the star. The mass
loss controls the termination of the AGB phase. When the envelope
becomes optically thin, the star leaves the AGB phase
\citep{lattanzio}. Shock waves can produce the necessary physical condition
(SiO abundance, kinetic temperature and density) for collisional
pumping of SiO maser \citep{elitzur}. Their positions is also consistent with
that of optically thick molecular layers responsible for radiative
pumping of these masers \citep{gray}.

The gas in the circumstellar envelope is largely molecular with a composition that
reflects the 
stellar chemistry. The molecules are located in different parts in the
envelope in shells of different radii and widths, depending on the
formation process and their resistance to shocks and ultraviolet
photons \citep{olofsson}.  

 The gas outflow of AGB star is normally considered to be  spherically
symmetric. Asymmetric structures are observed in  AGB stars, for
example in the radio photospheres of Mira, W Hya and R Leo  \citep{reid} and in
optical photosphere  \citep{karovska}.

R Cassiopeia is an oxygen-rich AGB star which is classified as an
M-type Mira-variable. Its optical brightness varies from magnitude
$+4.7$ to $+13.5$ with a period of 430 days and its mass is about 1.2
M$_{\odot}$. \citet{vlemmings} used astrometric  VLBI  to measure a distance
of $176^{+92}_{-45}$ pc with
a proper motion of (85.5 $\pm$0.8, 17.5 $\pm$0.7) mas yr$^{-1}$ in
R.\ A. and Dec., respectively. Various estimates of the stellar
velocity $V_{\star}$ appear in the literature; we tabulate these in
Table~\ref{t3} and adopt the mean value, +24 km s$^{-1}$, standard
deviation 2 km s$^{-1}$.

Because the maser emission is so bright, it allows the inner
circumstellar envelope to be resolved on sub-milliarcsecond scales.
The SiO masers are found in a region close to the star, within the
dust formation zone, located at a radius of about 2 to 4 stellar
radius (R$_{\star}$). They are observed as clumpy, partial rings
centred on the star; the masers fit a model of a hollow, expanding
ellipsoid as observed in TX Cam and U Her. The masing action needs
a long path through a relatively constant tangential velocity to
produce the necessary gain to be visible, this condition is often met
for a line tangent to the masing region with low turbulence because the
turbulence would reduce the path length at a given velocity limiting
the gain and hence the brightness \citep{diamond94}. 

The H$_2$O maser is
located in the inner parts of the circumstellar shell between $\approx
10$ to $20$ R$_\star$ where the temperature ranges from 300 K to 1000
K and densities are $10^7 - 10^9$ \,cm$^{-3}$ \citep{benson}. The OH
masers are found in the region furthest from the star as described by
\citet{cohen}.

\section{Observations and Data reduction}

The SiO masers around R Cas were observed as part of a more extensive
programme of VLBA\footnote{The VLBA (Very Long Baseline Array) is
operated by the the National Radio Astronomy Observatory, a facility
of the National Science Foundation operated under a cooperative
agreement by Associated Universities, Inc.} monitoring of other
stars. In total we have 23 epochs of data for R Cas. Data were
recorded at each VLBA antenna in dual-circular polarization in two 4
MHz windows, each digitally sampled at the full Nyquist rate of 8 Mbps
in 1-bit quantization. The lower spectral window was centred at a
fixed frequency corresponding to the v=1, J=1-0 SiO transition, at an
assumed rest frequency of 43.12207 GHz and a systemic velocity
V$_{\mathrm {LSR}}$ = +24 km s$^{-1}$. R Cas was observed for three
45-minutes periods evenly spread over the 8 hour duration of the
run. Adjacent to each R Cas observation, 5 minutes was spent observing
the continuum calibrator 0359+509 at the same frequency as R Cas. The
data were correlated at the VLBA correlator in Socorro, NM. The
correlator accumulation interval was set to 2.88 seconds. All
polarization correlation products (RR, RL, LR, LL) were formed. This
configuration produced auto- and cross-power spectra in each 4 MHz
baseband with a nominal velocity spacing of $\sim$ 0.2
km.s$^{-1}$. Table~\ref{t1} shows the epoch code, the date of
observation and the corresponding optical phase.

We reduced the data using the standard approach to VLBI
spectroscopy within the NRAO AIPS package
(http://www.aips.nrao.edu/cook.html). We processed the visibility 
data using the the semi-automated spectral-line polarization
calibration pipeline described by \citet{kemball95} and
further developed by \citet{kemball97}.  We summarise the main
steps here.  We started by editing the data. We derived corrections
for the phase-rate and the delay using the point-like calibration
source 0359+509. The Kitt Peak 
(KP) antenna has good stability in both pointing and receiver gain so
we used it as the reference antenna for all relevant stages at all
epochs.  We also scaled the KP autocorrelation spectra for 0359+509
with respect to the system temperature in order to derive the
amplitude scale for all cross-power data (see \citet{kemball09} for
details). We also used 0359+509 to correct for the phase response
across the bandpass. We applied the delay, rate, amplitude and
bandpass calibration to R Cas. 

At each epoch, we selected a reference group of strong channels from R
Cas and derived the residual rate corrections and then phase and
amplitude corrections, which were applied to all channels.  We took
care to preserve the low Stokes V signature during amplitude
calibration. We also corrected for parallactic angle
rotation. Additional polarization calibration and results will be
presented in a separate paper.  After all calibration, we produced
image cubes for R Cas at each epoch.  We imaged 56 channels around the
peak, each 0.217 km.s$^{-1}$ wide, image size $102.4\times102.4$ or
$204.8\times204.8$ mas, in order to enclose a greater extent than any
expected emission.  The naturally-weighted restoring beam varied from
epoch to epoch by a few mas and by a few degrees in position angle,
but the close similarities show that the visibility plane coverage was
consistent. The average beam size was
$(0.039\pm0.007)\times(0.020\pm0.008)$ mas$^2$ at average
position angle ($39\pm17$)$^\circ$. The thermal noise $\sigma_{\mathrm
  {rms}}$ in quiet channels was 30 mJy beam$^{-1}$.

We used the AIPS task {\sc sad} to fit 2-dimensional Gaussian
components to each clump of emission brighter than $5\sigma_{\mathrm
{rms}}$ in each channel, using the method of \citet{gonidakis}.  We grouped
components if their position in adjacent channels agreed within a
beamwidth $\sim$ 1 mas, rejecting isolated components. We term these
groups `features' and we found the error-weighted mean position and
flux-weighted mean velocity of each feature.  Since there was no
external phase-referencing, all positions were initially obtained
relative to the brightest component at each epoch. We aligned the
emission at different epochs following the method described by
\citet{cotton}. For each candidate position we found the flux-weighted
mean radius of all SiO maser components and then the inner and outer
radii which enclosed 90 percent of the total emission. The centre of
expansion is defined as the position giving the narrowest ring
enclosing 90 percent of the emission. We then inspected the changes in
component distribution relative to each centre. Some series of
components formed very distinctive patterns which could be used to
trace the emission from one epoch to the next. In most cases,
alignment using the fitted centres of expansion gave very small and
approximately symmetric shifts. However, at some epochs the maser
distribution was very lop-sided, the shells were very poorly defined
and we adjusted the apparent centres by eye to minimize the position
changes of distinctive features. The estimated radii are given in
Table~\ref{t2}. We estimated the errors arising from:
\newline (a) Signal-to-noise ratio based fitting errors from {\sc
sad};\newline (b) Difference between finding the centre by eye and the
automated fitting.

\clearpage

\begin{table}
\begin{center}
  \begin{tabular}{|c|c|c|} \hline
Epoch Code & Observing Date & Optical Phase $\phi$ \\ \hline

BD62A      & 1999-Sep-09   & 0.158 \\
BD62B      & 1999-Oct-15   & 0.241 \\ 
BD62C      & 1999-Nov-14   & 0.310 \\  
BD62D      & 1999-Dec-19   & 0.390 \\ 
BD62E      & 2000-Jan-15   & 0.452 \\  
BD62F      & 2000-Feb-14   & 0.521 \\ 
BD62G      & 2000-Mar-17   & 0.595 \\ 
BD62H      & 2000-Apr-21   & 0.675 \\ 
BD62I      & 2000-May-21   & 0.744 \\ 
BD69A      & 2000-Jun-22   & 0.818 \\ 
BD69B      & 2000-Jul-16   & 0.873 \\ 
BD69C      & 2000-Aug-21   & 0.956 \\ 
BD69D      & 2000-Sep-21   & 1.027 \\ 
BD69E      & 2000-Oct-19   & 1.091 \\ 
BD69F      & 2000-Nov-17   & 1.158 \\ 
BD69G      & 2000-Dec-20   & 1.234 \\ 
BD69H      & 2001-Jan-20   & 1.305 \\ 
BD69I      & 2001-Feb-19   & 1.374 \\ 
BD69J      & 2001-Mar-16   & 1.432 \\ 
BD69K      & 2001-Apr-15   & 1.501 \\ 
BD69L      & 2001-May-19   & 1.579 \\ 
BD69M      & 2001-Jun-24   & 1.662 \\ 
BD69N      & 2001-Aug-16   & 1.783 \\ \hline

  \end{tabular}

\caption{Observing dates and epochs.} \label{t1}
\end{center}

\end{table}

\begin{figure}
\begin{center}
\includegraphics[width=85mm,height=65mm ]{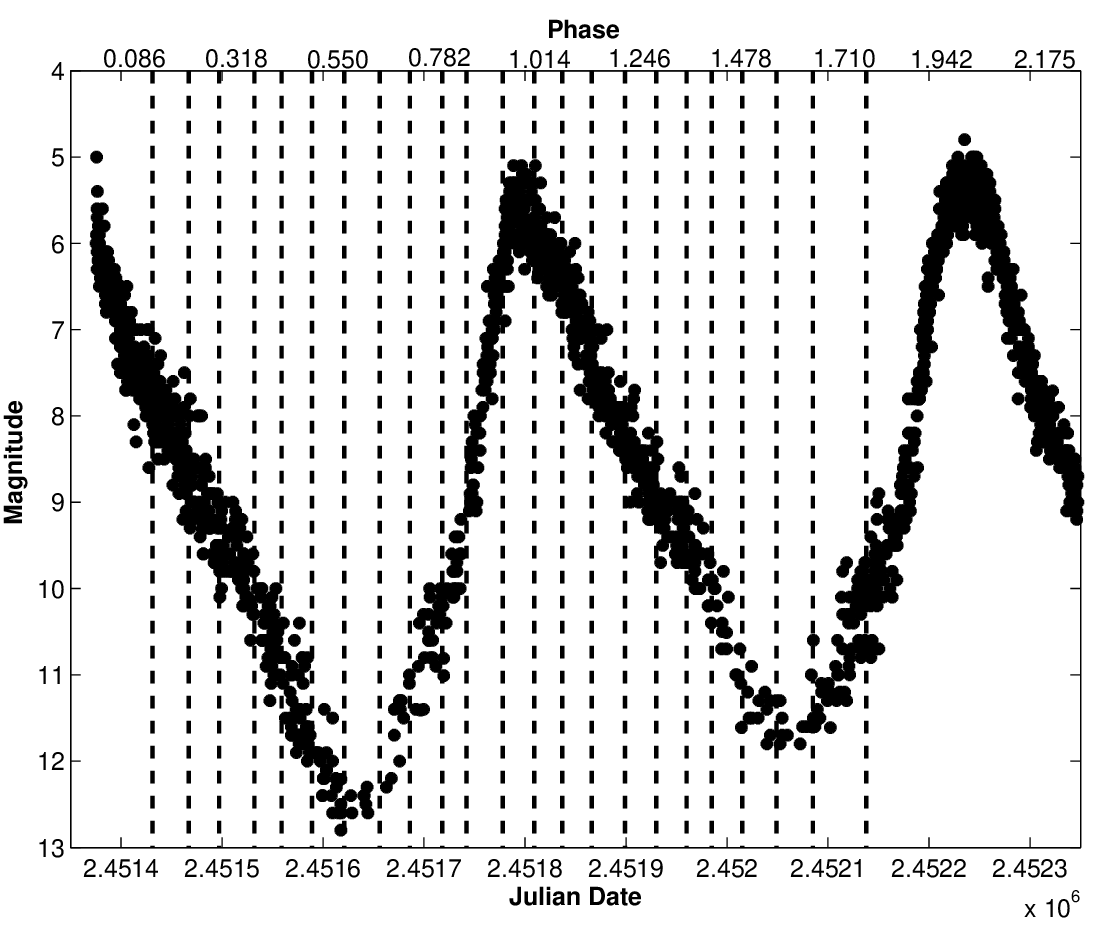}

\caption{The light curve of R Cas provided by the AAVSO. The
  dashed vertical lines indicate the dates of our observations.}
\label{f1}
\end{center}
\end{figure}

\begin{figure}
\begin{center}
 \includegraphics[ width=95mm, height=55mm]{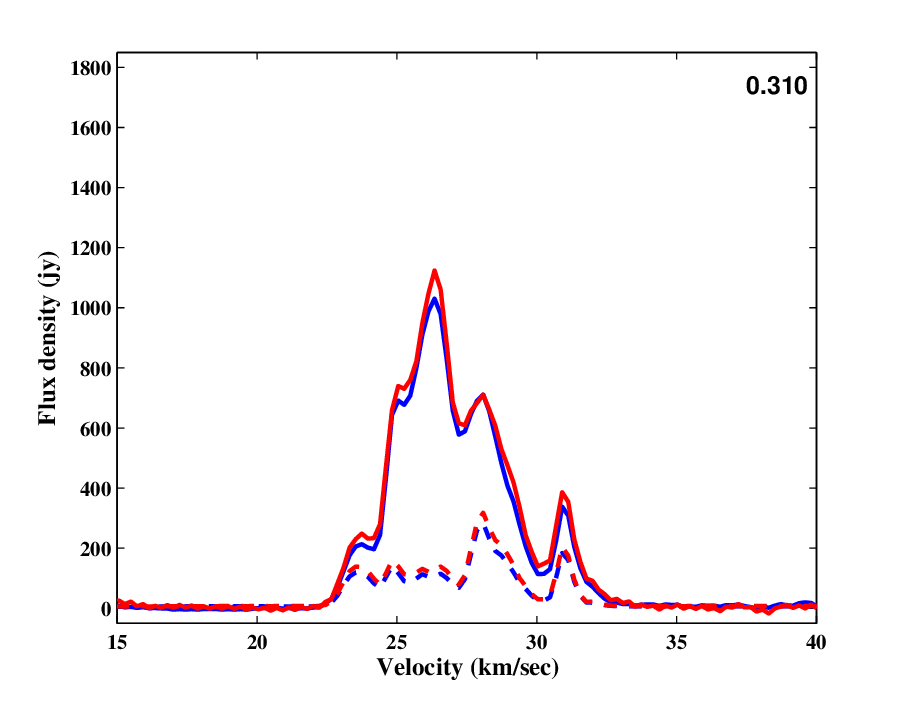}
  \caption{The flux density detected by the KP antenna (solid line)
  and by the shortest baseline (LA -- PT) (dashed line), for the two
  hands of polarization (red=right-hand, blue=left-hand).}
\end{center}
\label{f2}
\end{figure}

\begin{table}
\begin{center}
\begin{tabular}{|c|c|c|} \hline
      Epoch & Radius (mas) & error (mas)\\ \hline 
      BD62A & 25.46        & 0.67       \\ 
      BD62B & 27.09        & 0.05       \\
      BD62C & 27.08        & 1.77       \\
      BD62D & 26.05        & 1.22       \\
      BD62E & 27.10        & 1.59       \\
      BD62F & 28.43        & 1.98       \\
      BD62G & 27.41        & 2.11       \\
      BD62H & 28.09        & 0.3        \\
      BD62I & 28.49        & 0.53       \\
      BD69A & 27.29        & 1.64       \\
      BD69B & 27.71        & 0.37       \\
      BD69C & 28.00        & 1.5        \\
      BD69D & 27.98        & 0.16       \\
      BD69E & 27.50        & 0.5        \\
      BD69F & 28.15        & 1.71       \\
      BD69G & 28.52        & 2.5        \\
      BD69H & 29.16        & 0.26       \\
      BD69I & 27.15        & 0.17       \\
      BD69J & 26.83        & 0.63       \\
      BD69K & 25.11        & 3.11       \\
      BD69L & 23.96        & 1.22       \\
      BD69M & 25.87        & 3.83       \\
      BD69N & 26.82        & 1.68       \\ \hline
\end{tabular}
\caption{Estimated radii of the SiO maser shell at each epoch. } \label{t2}
\end{center}
\end{table}

\section{Results}
The optical light curve of R Cas provided by the AAVSO\footnote{We
acknowledge with thanks the variable star observations from the
AAVSO (American Association of Variable Star Observers)
International Database contributed by observers worldwide and used
in this research.}, is shown in Fig.~\ref{f1}. The dashed 
lines indicate  the VLBA observational epochs that span the time
interval from 1999 September 09  to 2001 August 16, corresponding to
an optical phase interval of $\phi = 0.158$ to $\phi = 1.783$. 
Fig.~\ref{f2} shows the flux density in Jansky of SiO maser emission
toward R Cas in both RR and LL polarizations. In this figure, the
solid line indicates the single-dish flux density for the Kitt Peak
antenna, and the dashed line, the flux density for the shortest VLBA
baseline PT-LA. It clear from the 
figure that the shortest baseline of the VLBA detected about  27
percent  of the single dish SiO flux.  Comparisons for all epochs are
shown in Figs. B1 -- B3.

The gross morphology of the SiO maser emission toward R Cas in each
individual epoch is shown in Fig.~\ref{f3} and
Fig.~\ref{f4}. Predominantly, the maser emission is confined to a
narrow projected ring in phases $\phi = 0.241 - 1.374$. The first
epoch, $\phi = 0.158$, shows an additional, outer arc; a few outlying
masers are also seen at $\phi = 0.744 - 0.873$.  From $\phi = 1.432 -
1.783$, the maser emission has an irregular shape, and the maser spots
are very faint. The shell is dominated by an Eastern arc during the
first stellar period and by a Western arc during the second
period. Fig.~\ref{f5} shows the SiO maser features for all 23 epochs,
superimposed on one plot. This shows that
the SiO emission is predominantly situated in a ring that is poorly
filled at the north and south. This ring has a radius of $1.6-2
R_\star$ depending on the obsering epoch.

Fig.~\ref{f6} shows the observed ring radii (tabulated in
Table.~\ref{t2}) plotted as a function of stellar phase. The error bars
on the points indicate the ring thickness, computed as the standard
deviation in the observed ring radius. We fitted a parabolic function
(solid line) to the data in Fig.~\ref{f6} that shows the increase in
the ring diameter over the phase range $\phi \sim 0.1$ to $\phi \sim
1.3$ \citep{diamond03}. The projected shell starts shrinking beyond the
optical phase $\sim 1.3$.

\section{Discussion}

\subsection{Morphology}
\begin{figure}
\begin{center}
\includegraphics[width=78mm,height=69mm ]{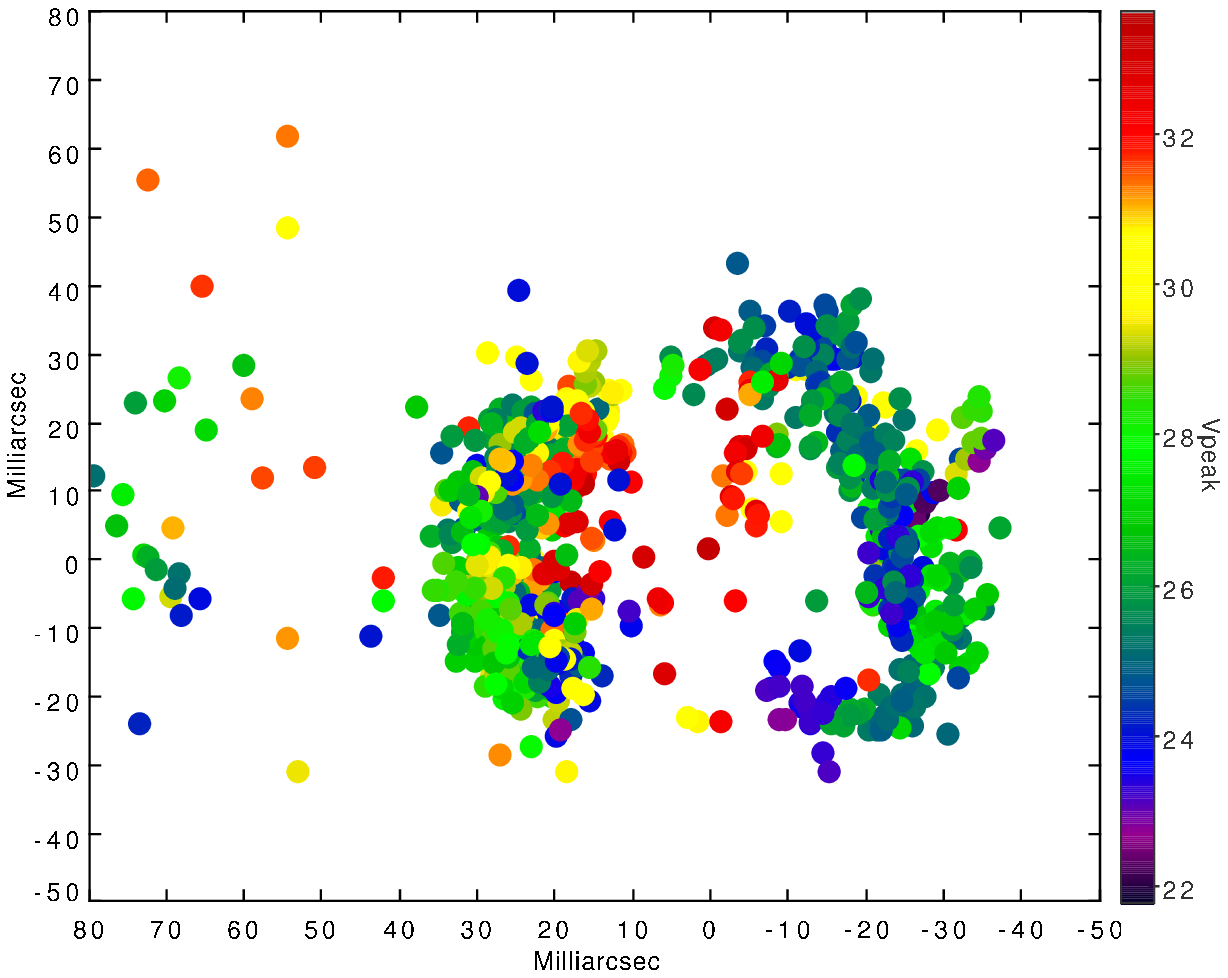}
\caption{SiO maser features observed at all 23 epochs.}
\label{f5}
\end{center}
\end{figure}

 The ring-like appearance of the SiO masers (Fig.~\ref{f5}) is thought
to arise from a spherical shell, wherein strong radial acceleration or
deceleration causes predominantly tangential amplification. The
remnant of an outer, presumably older, shell is seen in the East in
Fig.~\ref{f3} ($\phi = 0.158$) and Fig.~\ref{f4}. At  this first epoch
and at $\phi \ge 1.234$, some faint emission is seen within 20 mas of
the centre of expansion. This could be due to the emergence
of a new shell and, indeed, the shell radius shrinks at $\phi\ge1.374$
(discussed further in Section~\ref{sec:diameter}). Alternatively, the central
masers could be radially beamed from parcels of gas temporarily moving
at a fairly steady velocity, in between
acceleration and deceleration. Most of the inner masers are
red-shifted, suggesting infall, since they lie along the line of sight
to the star, which would be optically thick to maser emission from the
back of the shell.

\subsection{Maser shell asymmetry}

\begin{figure}
\begin{center}
\includegraphics[width=100mm,height=218mm ]{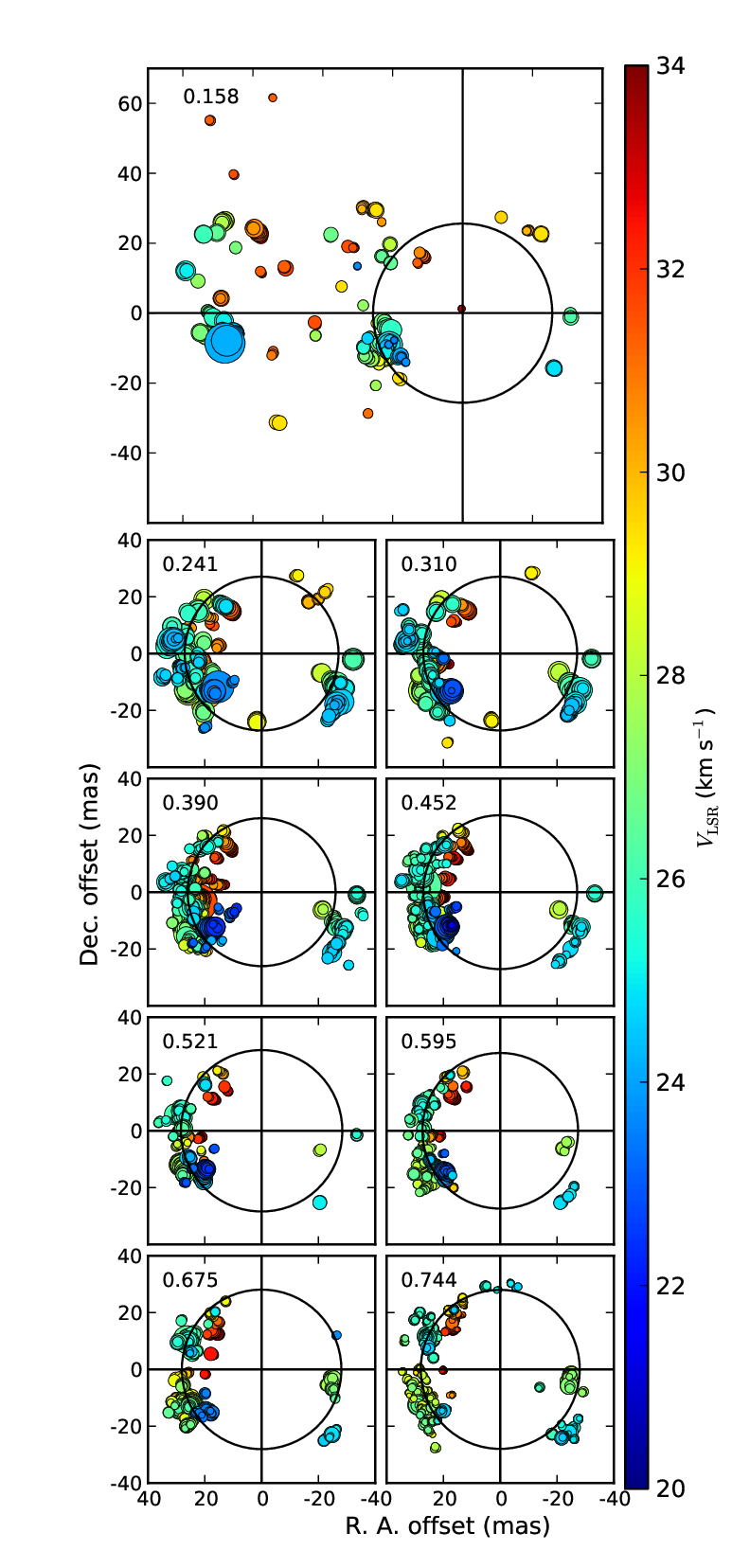}
\caption{Total intensity images [Epochs (1-9)]. The symbol size is
  proportional to the square root of the flux density of each
  component, the large circles represent the  mean maser shell.}
\label{f3}
\end{center}
\end{figure}

\begin{figure}
\begin{center} 
\includegraphics[width=185mm,height=210mm ]{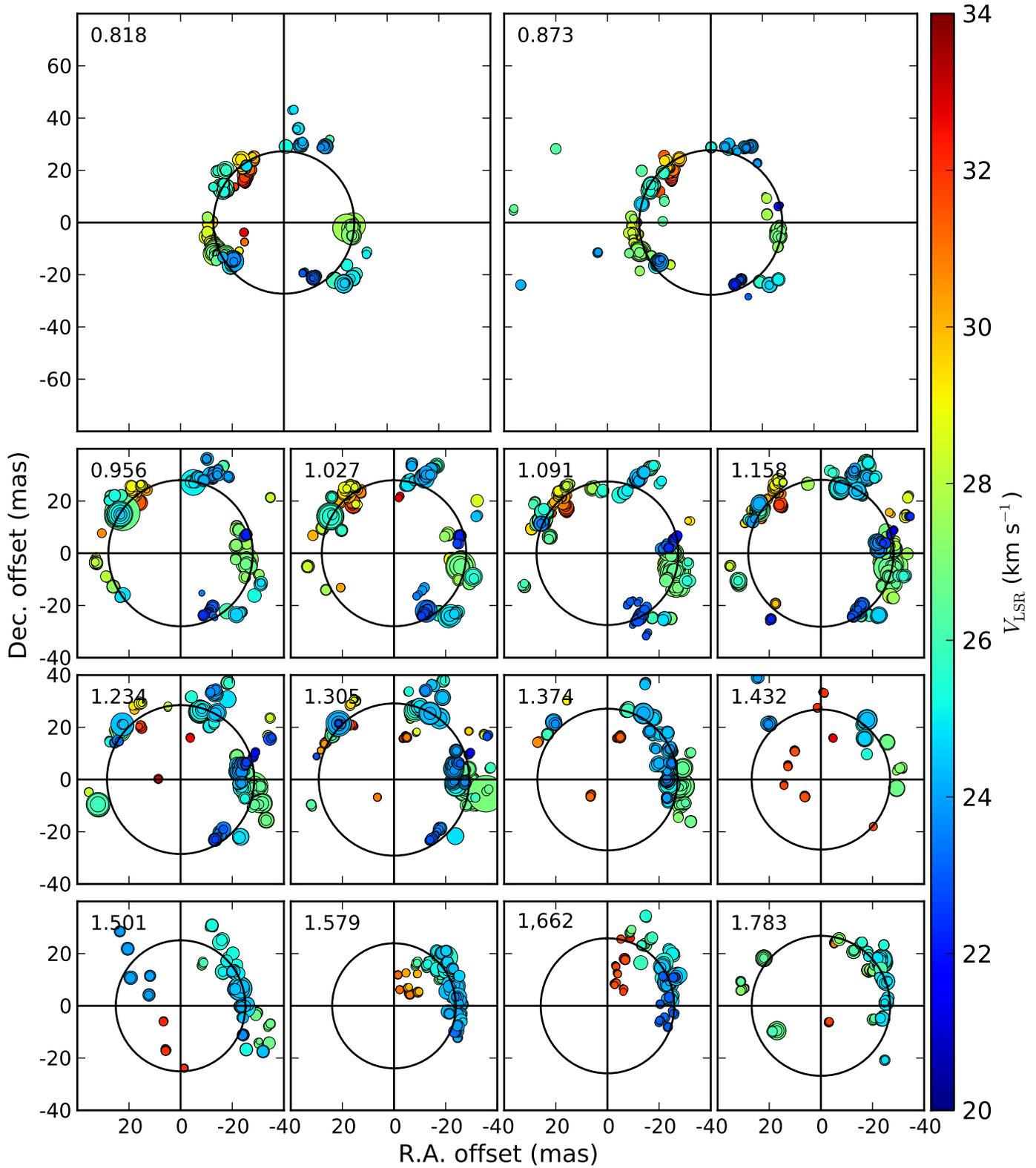} 
\caption{Same as Fig.~\ref{f3} [Epochs (10-23)].}
\label{f4}
\end{center}
\end{figure}

\begin{figure}
\begin{center}
\includegraphics[ width=95mm, height=55mm]{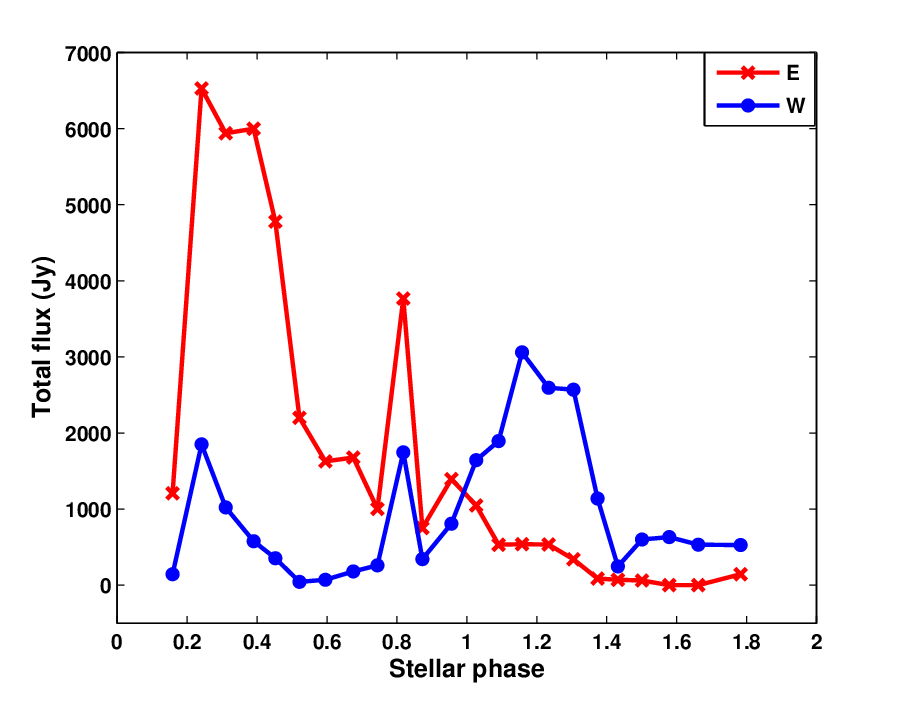}
\caption{The total East- (Red) and West- (Blue) flux density for 23 epochs. }
\label{f7}
\end{center}
\end{figure}

Figs.~\ref{f3} and~\ref{f4} show that the SiO emission is dominated
by an Eastern arc during the first stellar cycle, whilst the Western
side of the shell dominates for the remaining epochs. More than 3/4
of the emission comes from the East at $\phi=0.158 - 0.744$ and more
than 3/4 comes from the West at  $\phi=1.091 - 1.783$ as it is seen in
Fig.~\ref{f7}.  A similar 
dominance by one (Western) arc was seen by \citet{phillips01} in two
epochs of observation of the 86-GHz SiO masers in 1998. There are
several possible explanations.

Firstly, if conditions in one hemisphere of a spherical SiO maser shell are
more favourable for masing, the change in E--W asymmetry from phase to
phase could be due to bulk rotation, away from us in the E and towards
us in the W, about an N--S axis.  This would require an equatorial
velocity of 10--20 km.s$^{-1}$ which is about twice the observed line
of sight velocities, but this could be due to turbulence and selective
maser amplification in our direction. However, it is hard to conceive
a physical mechanism for spherical solid-body rotation, although a
flared disc might be possible.

Secondly, the star may eject mass into arcs, as has
been seen in IRC+10216 \citep{murakawa}, which could mean that at some
cycles, mass ejections were concentrated along the line of sight. In
such cases we would see little or no SiO maser emission.

Thirdly, stellar activity may disrupt masing close to the
star. \citet{rudnitskij08} report that R Cas is known to display
H$\alpha$ flares at intervals of several years (also seen from other
Miras). This is the most probable explanation although the second
factor could play a part.

\subsection{Maser shell diameter}
\label{sec:diameter}
Fig.~\ref{f6} shows the variation of the ring diameter for 23 epochs
as a function of stellar phase.   It is evident that expansion is the
dominant overall kinematic behaviour between optical phases $\phi \sim
0.158 $ to $\phi \sim 1.3$. Thereafter, the diameter decreases until a
phase of $\phi \sim 1.58$. This may be due to the disappearance of
outer maser features and/or infall under the influence of gravity. At
$\phi > 1.58$ the diameter increases again, suggesting the impact of a
new stellar pulsation. Similar behaviour was seen in TX Cam
\citep{diamond03}.

  We found the gravitational acceleration ($g_{\mathrm {SiO}}$) was
$\sim3.27 \times 10^{-7}$ km s$^{-2}$ at the midpoint of SiO maser
region ($\sim$ 26.56 mas) for a star mass of 1.2 M$_{\odot}$. By
adopting a distance to R Cas of 176 pc \citep{vlemmings}, we found that the
required infall time for a parcel of gas across a maser zone of size
$\sim$12 mas is $\sim500$ days. This estimate has an uncertainty of
$>20$ percent, mainly due to the uncertainty in the distance to R Cas,
but is comparable to, or greater than, the stellar pulsation period of
431 days. \citet{diamond03} found a similar relationship for TX Cam.
We also used our estimate of $g_{\mathrm {SiO}}$ to calculate that, if
a parcel of gas was stationary at $\phi$ = 1.308, it would attain a
velocity of 14 km s$^{-1}$ at $\phi$ = 1.579, which is almost equal to
the velocity of 13.7 km s$^{-1}$ required to reach the observed
minimum radius.

These results indicate that complex kinematics exist in
the SiO maser region. The gas is subject to two main forces; 
outward pressure due to stellar pulsations and the inward force of
gravity. Fig.~\ref{f6} shows that expansion appears to be a slower
process than infall over the same distance, consistent with a variable
outward force but a more constant inward force.
When the stellar pulsation pressure is weakest, at the outer
edge of the SiO maser shell, the parcel of gas would
start to fall back towards the star under its gravitational pull.  If 
the parcel does not reach the inner edge of the shell before the next
pulsation, it will then be pushed out futher. These results, like
those for TX Cam, are a step towards quantative verification of
pulsation models like \citet{bowen}.

\subsection{Maser models}

 We make some simple comparisons between the observed maser structure
 and the model of \citet{gray}; a more detailed analysis will
 be presented in a future paper. The model deals with tangentially
 beamed emission, but although emission is seen along the line of
 sight to the star at some epochs, this is too faint to influence
 comparison of the models with the average velocity profiles or radii.
 This model predicts that the diameter of the maser ring should be
 about twice the size of the stellar photosphere. The optical stellar
 diameter of R Cas is $25.3\pm3.3$ mas at $\phi=0.93$ \citep{weigelt}. We
 found that the 
 SiO maser mean angular diameter, averaged over 23 epochs, is $\sim54$
 mas which is double the photospheric diameter within the
 uncertainties, showing that the model is consistent with our data.

\begin{figure}
\begin{center}
 \includegraphics[ width=80mm, height=60mm]{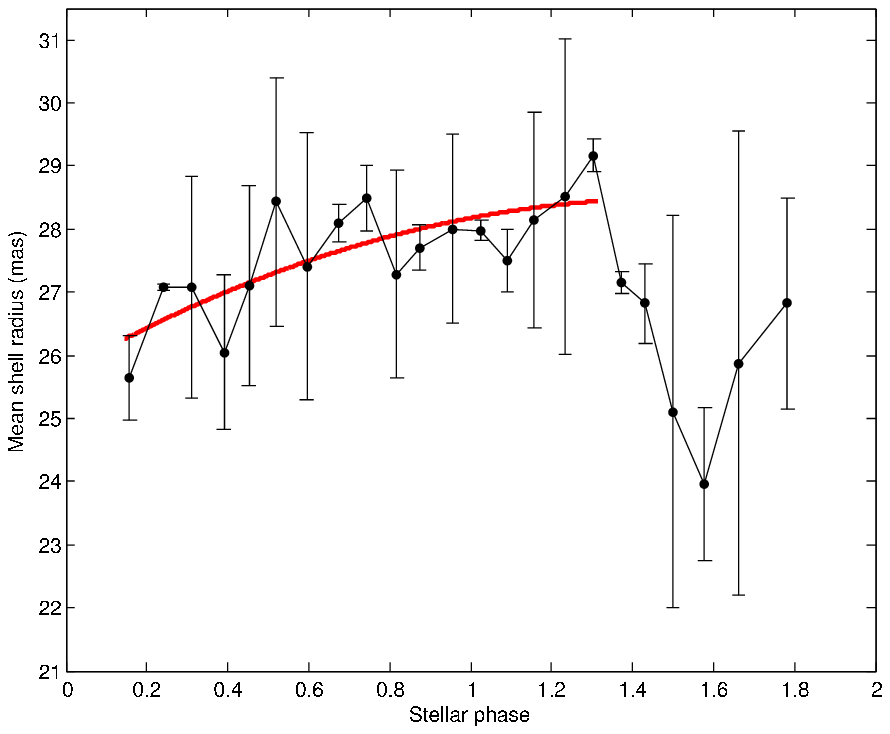}
  \caption{The variation of the SiO maser ring radius with the stellar
  phase.}
\label{f6}
\end{center}
\end{figure}

\begin{figure}
\begin{center}
  \includegraphics[width=80mm, height=55mm]{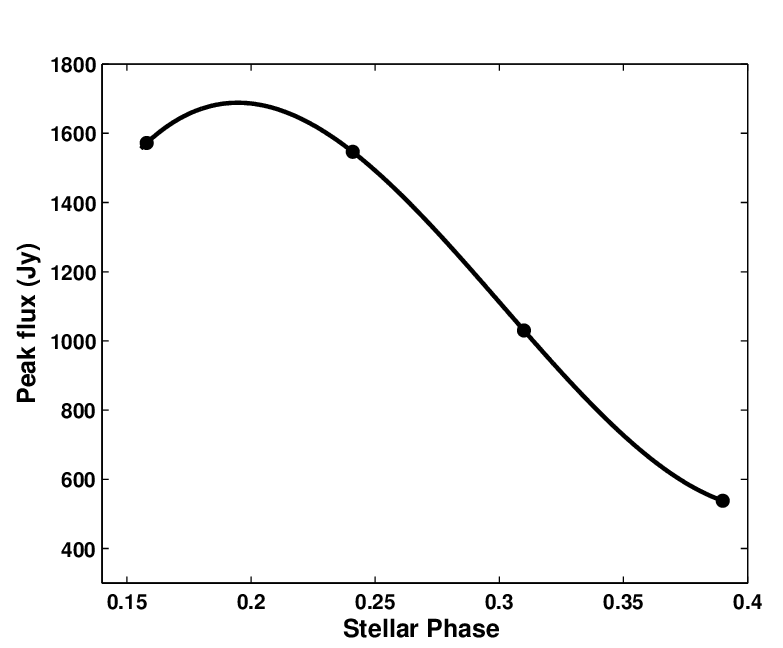}
\caption{The peak SiO maser flux density as a function of the stellar phase
  over the interval described by the model of \citet{gray}.}
\label{f8}
\end{center}
\end{figure}

This model also predicts that the ring size increases from stellar phase
0.1 to 0.25 and then decreases.  Fig.~\ref{f6} shows that, during the first
cycle, the radius continues increasing past phase 0.25.  However, in the
second cycle, the radius decreases sharply at the predicted
phase.. The model predicts that the 
 maser weakens between phases $0.25$ and $0.4$. Fig.~\ref{f8} shows
 that the flux density peaks at $\phi \sim 0.2$ and the then
 decreases over the same phase range which suggests good consistency
 between our observations and the model.

\begin{figure}
\begin{center}
 \includegraphics[width=90mm, height=47mm]{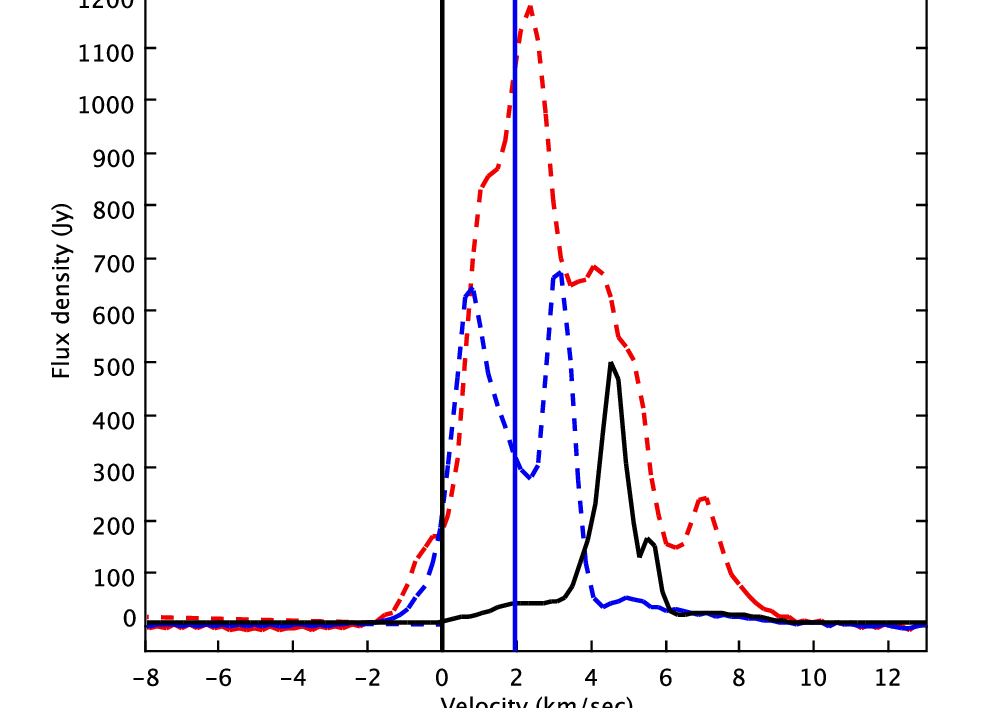}
\caption{Total intensity SiO maser velocity profiles measured by a
  single dish. The red and blue dashed curves show the emission
  averaged over the first and second cycles, respectively.  The black
  solid curve represents the model. The black vertical line is the
  stellar velocity with an uncertainty ($\pm$2 km.s$^{-1}$) and the
  blue vertical line is the flux-weighted mean velocity for all
  epochs. }
\label{f9}
\end{center}
\end{figure}

 Another prediction of the model is that the average velocity of the
 transition studied here moves to higher, redshifted velocities, from
 phase 0.1 to 0.4. We tested this using the single dish
 (autocorrelation) observations shown in Figs. B1 -- B3, in order to
 ensure that the full spectra were measured.  The red and blue, dashed
 lines in Fig.~\ref{f9} show the velocity structure averaged over this
 range of phase for the first and second cycles, respectively. The
 black vertical line indicates the stellar velocity (with an
 uncertainty of 2 km s$^{-1}$). The blue vertical line marks the flux
 weighted mean velocity for all epochs. The black, solid line is the
 model. This figure shows that the peak flux density is red-shifted in
 both the present R Cas data for the first cycle and in the model. The
 separation in the velocities of the peaks in the model and R Cas
 spectra is just 1.25 standard deviation in the R Cas data.The second
 cycle is not consistent with the model. 

\begin{figure}
\begin{center}
  \includegraphics[width=85mm, height=50mm]{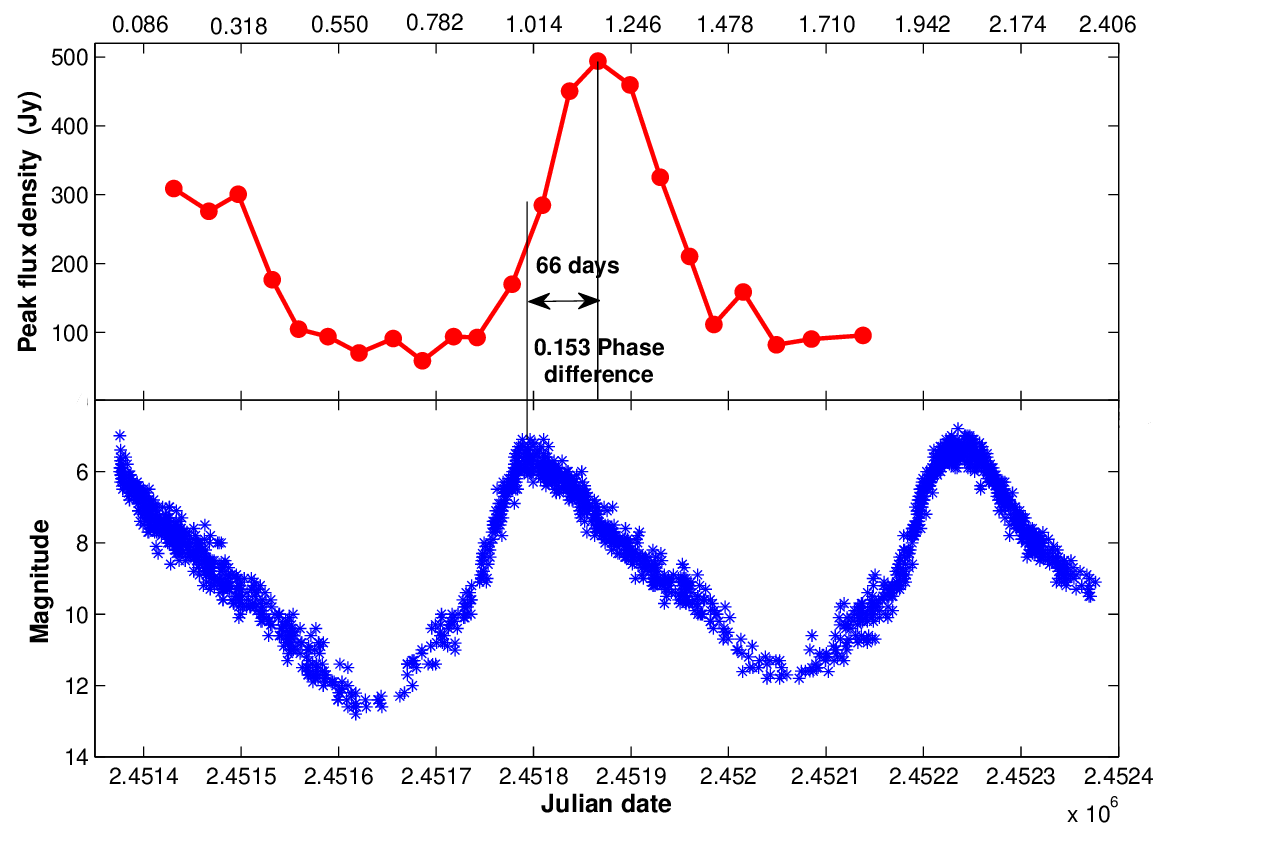}

\caption{Phase difference  between the maser light curve (upper frame) and
  the optical light curve (lower frame).}
\label{f10}
\end{center}
\end{figure}

 Fig.~\ref{f10} shows that the SiO maser variabilty in R Cas has approximately
 the same period as its optical light curve with a phase lag
 $\approx$ 0.153 (66 days) which is consistent with the predictions  of
 \citet{pardo}.

At this point, we compare the observational data of R Cas with a
computational model of an AGB star. This model combines hydrodynamic
solutions with a maser radiative transfer code as described in \citet{gray},
apart from the exceptions discussed below.  

\begin{figure}
\begin{center}
 \includegraphics[ width=80mm, height=45mm]{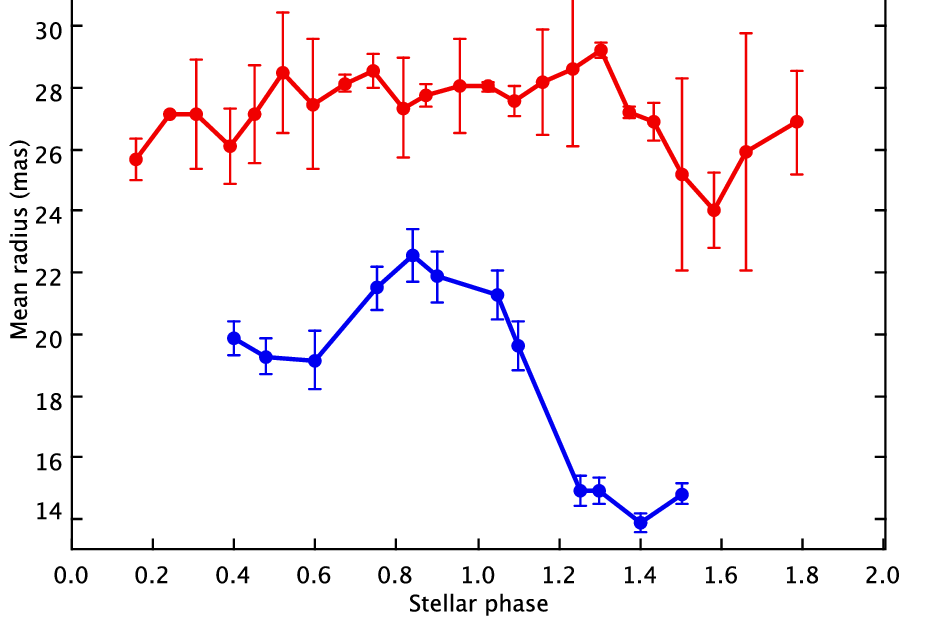}
 \caption{A comparison between the behaviour of the radius in the
 model (blue) and our data (red).}
\label{f11}
\end{center}
\end{figure}

The \citet{gray} model was extended to sample 12 stellar phases between 
$\phi=0.4$ to $\phi=1.5$, so that the model
pulsation passes through an optical maximum. We  note  that the model
star is not specifically  designed to represent R Cas. In \citet{gray},
the 4 phase samples were modelled as separate snapshots,  
unrelated to each other in the sense of positional evolution of the
masing objects. By contrast, the current model evolves the maser
distribution by taking the output conditions of one sample as the input
conditions for the next, as in \citet{humphreys}. Therefore the
maser spots have random positions only for the first sample (at phase
0.4 in the first cycle). We computed the inner, outer and mean radius
for each model epoch, defined in the same way as for the
observational data. The mean radius of the maser shell is plotted for R Cas, in
Fig.~\ref{f11}. It is evident that the model maser shell is substantially
smaller than R Cas. However, a significant decrease in mean radius is apparent
following optical maximum in both model and R Cas. This is the most
statistically significant radius change in the observational
data. This shrinkage is delayed in the R Cas data, with respect to the
model, by about 0.2 stellar periods. On inspection of the model
spectra during the radius drop, the peak of the model 43-GHz spectrum
shifts to the red by  $\sim$7 km.s$^{-1}$.

\clearpage

\clearpage
This is consistent with the observed
reddening  of the  R Cas emission during its contraction. We also
inspected the model velocity at mean radius $\bar{R}$, and noted that
the velocity was closer to infall at the $\bar{R}$ during ring
contraction than at optical maximum.

\section{Conclusions}
The SiO maser morphology in R Cas for the transition v = 1, J=1$\to$ 0
is described as a rarely completed ring structure, and the ring is
dominated  by an Eastern arc in the first stellar period and by a
Western arc in the second stellar period. The result are compared with
a model produced by \citet{gray} which predicts most of observed
features in our data.The infall time towards the star is greater than
the pulsation period with a factor of $\sim$ 1.2. The maser light
curve behavior is similar to the optical light curve but with phase lag of
66 days.

\section*{Acknowledgments}

Many thanks to  VLBA for providing the data used in this paper and
many thanks to AAVSO for providing the the optical light
curve.The authors acknowledge the use of the UCL Legion
High Performance Computing Facility, and associated
support services, in the completion of this work. K.A.A. would like to
extend  his thanks to the Iraqi 
government for giving him   the opportunity to do the PhD study at the
University of Manchester and he thanks his friend Qusay Al-Zamil  for some
helpful  suggestions  that improved the presentation of the paper. We
would like to extend our thanks to the referee for his helpful suggestions.

\appendix 

\section{Estimates of the stellar velocity}

Various estimates of the stellar
velocity $V_{\star}$ appear in the literature; we tabulate these in
Table~\ref{t3} and adopt the mean value, +24 km s$^{-1}$, standard
deviation 2 km s$^{-1}$.
\begin{table*}
\begin{center}
\begin{tabular}{|c|c|} \hline
    V$_{\mathrm {LSR}}$ (\,km. \,s$^{-1}$) & Reference\\ \hline 
      20.4        & \citet{jewell} \\ 
      24          & \citet{colomer} \\
      26          & \citet{vlemmings} \\
      25.4        & \citet{boboltz} \\
      24.9        & \citet{matthews} \\ \hline
\end{tabular}
\caption{Stellar velocity.} \label{t3}
\end{center}
\end{table*}

\section {Plots of flux densities from a single dish antenna and from
  the shortest baseline for dual-hand  polarization} 
\begin{figure*}
\begin{center}
 \includegraphics[width=140mm, height=200mm]{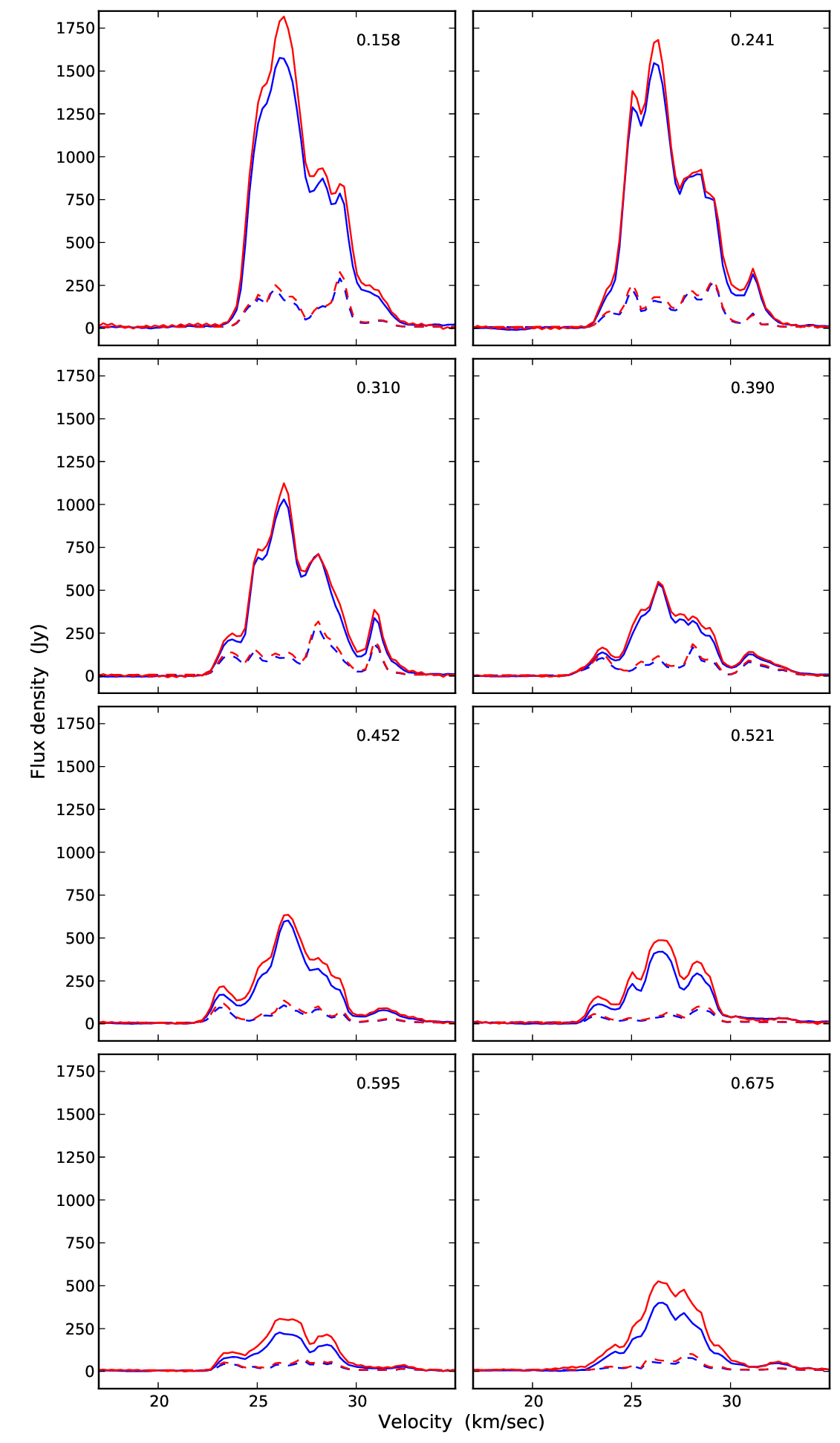}  
  \caption{The flux density for antenna KP (solid line)  and shortest
  baseline (LA -- PT) (dashed line) for the two hands of polarization
  (red=right-hand, blue=left-hand) for epochs 1 -- 8.}.  \label{ap1} 
\end{center}
\end{figure*}

\begin{figure*}
\begin{center}
 \includegraphics[width=140mm, height=200mm]{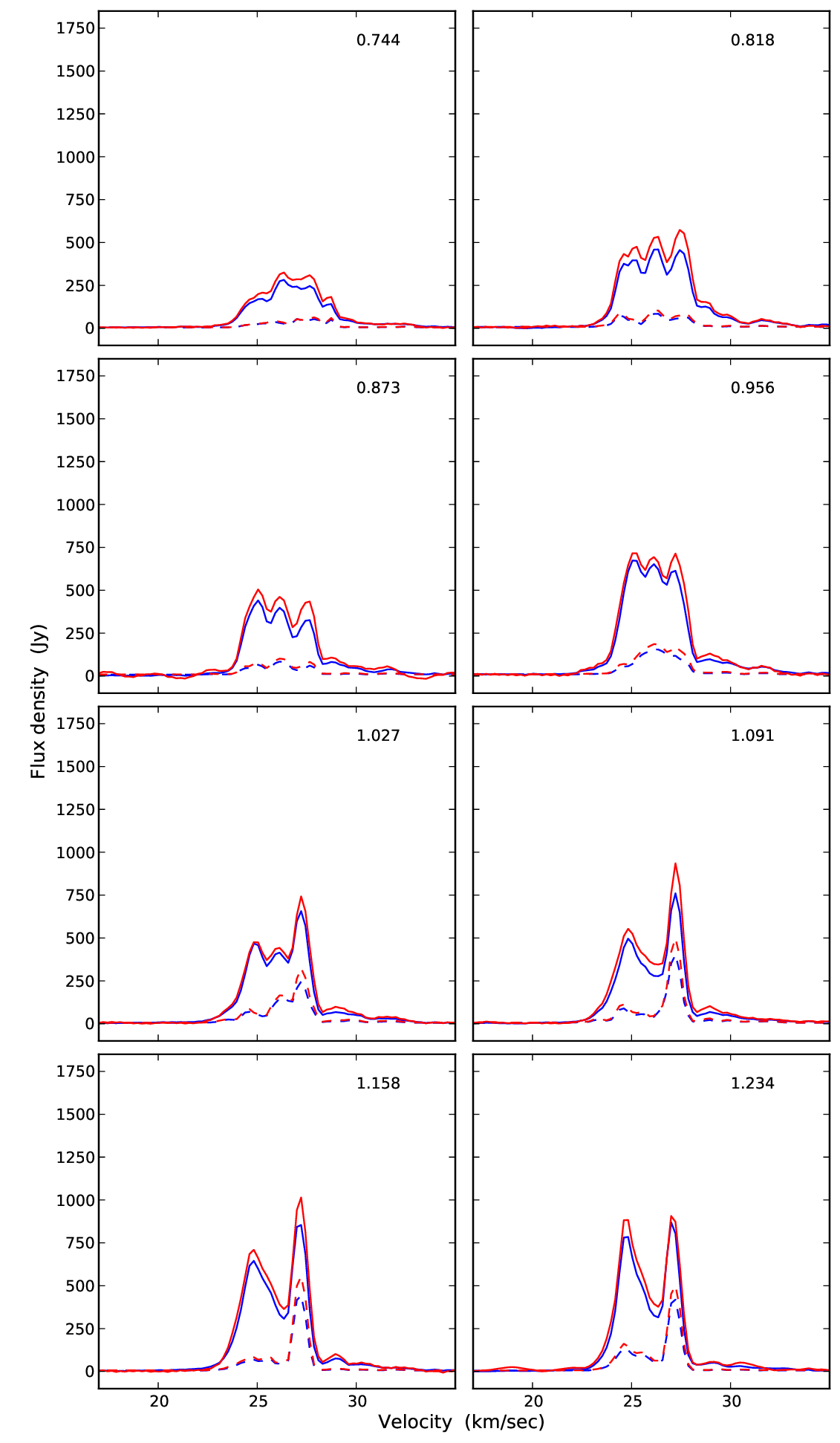}
  \caption{Same as Fig. B1 for epochs 9 --16.} \label{ap2}
\end{center}
\end{figure*}

\begin{figure*}
\begin{center}
 \includegraphics[width=140mm, height=200mm]{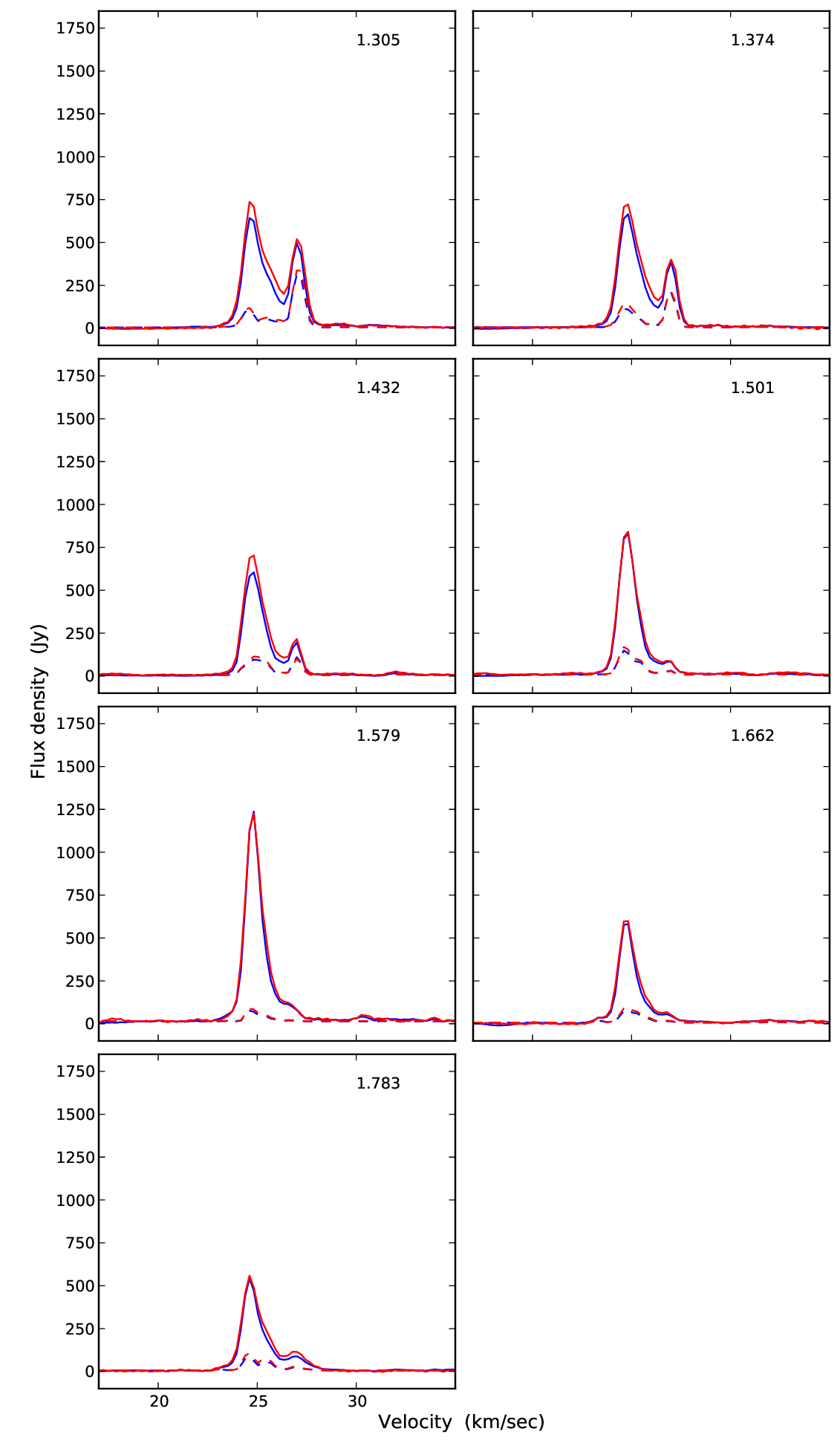}
 \caption{Same as Fig. B1 for epochs 17 -- 23.} \label{ap3}
\end{center}
\end{figure*}

Figs.~{\ref{ap1}, \ref{ap2}, \ref{ap3}} compare the flux density from a
single dish  Kitt Peak (KP) (solid line) and from the shortest baseline
between Los Alamos and Pie Town (LA -- PT) (dashed line) for the two
hands of polarization (red = right hand, blue = left hand). The
number on each plot indicates the stellar phase.The apparent
  offset between L and R is probably a residual error.
\clearpage

%\bsp

\label{lastpage}

\end{document}